\documentclass[11pt]{article}
\usepackage{graphicx}
\begin{document}
~~~\\
\begin{center}
{\bf Variational Principle for Mixed Classical-Quantum Systems \\[1cm]}
{ M. Grigorescu \\[3cm]  }
\end{center}
\noindent
$\underline{~~~~~~~~~~~~~~~~~~~~~~~~~~~~~~~~~~~~~~~~~~~~~~~~~~~~~~~~
~~~~~~~~~~~~~~~~~~~~~~~~~~~~~~~~~~~~~~~~~~~}$ \\[.3cm]
An extended variational principle providing the equations of motion for a system consisting of interacting classical, quasiclassical and quantum components is presented, and applied to the model of bilinear coupling. The relevant dynamical variables are expressed in the form of a quantum state vector which includes the action of the classical subsystem in its phase factor. It is shown that the statistical ensemble of Brownian state vectors for a quantum particle in a classical thermal environment can be described by a density matrix evolving according to a nonlinear quantum Fokker-Planck equation. Exact solutions of this equation are obtained for a two-level system in the limit of high temperatures, considering both stationary and nonstationary initial states. A treatment of the common time shared by the quantum system and its classical environment, as a collective variable rather than as a parameter, is presented in the Appendix. \\ 
$\underline{~~~~~~~~~~~~~~~~~~~~~~~~~~~~~~~~~~~~~~~~~~~~~~~~~~~~~~~~
~~~~~~~~~~~~~~~~~~~~~~~~~~~~~~~~~~~~~~~~~~~}$ 

\newpage
{\bf I. Introduction}  \\

The time-dependent variational principle (TDVP) provides the equations of motion for dynamical systems evolving on trajectories which are the critical points of an action functional, or the geodesics of a suitable metric \cite{am}. \\ \indent 
In classical mechanics the action is a functional defined on the infinite dimensional space of the trajectories  in the velocity, or momentum phase space, and locally the dynamics is expressed by the Euler-Lagrange, respectively Hamilton equations. The classical action remains significant also for the quantum dynamics. For instance, by the Bohr-Sommerfeld integrality constraints it can provide the exact energies of the stationary states, while in the Gamow formula it describes the stationary tunneling rates. In general, the transition amplitudes can be related either to the classical action, in the path-integrals formalism \cite{feyn}, or to a specific "quantum action" \cite{kroger}.
\\ \indent
The evolution of the state vectors in quantum mechanics is expressed essentially by the time-dependent Schr\"odinger equation (TDSE). This can be written as a Hamilton system of equations, and therefore it can be derived by the TDVP with an action functional (the "TDSE action") depending on trajectories in Hilbert space \cite{kram}. The TDVP can be applied to this action not only to recover TDSE, but also to obtain quasiclassical approximations, by constraining the variations to suitable manifolds of trial states. For example, in the case of a many-fermion system such variations within the manifold of Slater determinants provide the time-dependent Hartree-Fock equations \cite{rowe}. 
 \\ \indent
Dynamical systems in which a classical, deterministic part, interacts with a quantum part, have been a challenge since the early days of the quantum theory.  The presumed strong coupling switched on at the time of a measurement was taken into account by turning TDSE into an instantaneous collapse of the wave function. However, TDSE was successfully used to describe slow, continuous couplings, such as in the Born-Oppenheimer approximation \cite{boa}. Moreover, in the context of semiclassical gravity, Hamilton-Heisenberg \cite{aand}, or Hamilton-Schr\"odinger \cite{icqe} equations coupling quantum fields to classical metric have been derived using TDVP's with suitable action functionals. Though, such theories have limitations, as a system with negative energy density could generate fluctuating gravitational fields, producing a Brownian-like diffusion of the test particles \cite{ford}.
\\ \indent
Quantum systems interacting continuously with a non-deterministic, thermal environment, appear in the treatment of relaxation phenomena. The methods applied range from perturbative rate equations for the occupation probabilities \cite{vhove} to the quantum Fokker-Planck equation \cite{fpl}. For a quantum Brownian particle, by analogy with the classical equations of motion have been proposed quantum Langevin equations, in the Schr\"odinger \cite{kost, raz} or Heisenberg representations \cite{ford1}.  
\\ \indent  
The analogy with the classical description of the thermalization process can be used not only at the level of the phenomenological equations of motion, but also in the TDVP which provides the microscopic dynamics. Such a variational approach can be formulated by defining the action functional as the sum between the classical action and a TDSE action in which the Hamilton operator depends on the classical variables \cite{mg1, pha}. It was shown that in the presence of a classical thermal environment the quantum dynamics is described by a nonlinear Schr\"odinger-Langevin equation \cite{mg1, pha}, successfully applied to calculate heating-assisted atomic tunneling rates \cite{cjp}.  
\\ \indent
In this work the variational approach to the mixed classical-quantum dynamics is extended by several elements, concerning both the formalism and its applications. In Section II the TDVP is expressed in terms of a single action functional, the TDSE action, including the classical action in the phase factor of the state vectors. To cover also the intermediate case when some degrees of freedom are described by coherent states, the Hilbert space is factorized in a linear subspace and a "trial manifold". This framework is used in Appendix to show that time in quantum mechanics can be treated as a dynamical variable of an extended Hilbert space, rather than as a parameter.  \\ \indent
The proposed formalism is applied in Section III to a quantum particle interacting with a classical environment  by the "bilinear coupling" model of the transition-rates theory. It is shown that at finite temperature the particle can be described by a nonlinear Fokker-Planck equation for the average density matrix. This equation is solved in Section IV for a quantum two-state system, considering both stationary and nonstationary initial conditions. When friction is negligible, the solutions can be obtained in analytic form, showing explicitly the decoherence effect of the noise. The nonlinear friction term alone leads to a form of spontaneous decay, studied at zero temperature. Conclusions are summarized in Section V.  \\

{\bf II. The variational principle for classical-quantum systems} \\

Let us consider a classical system described by the momentum phase-space coordinates ${\cal C} \equiv (q^1,q^2,...,q^n,p_1,p_2,...,p_n)$ on the symplectic manifold $M_{cl}$, and a quantum system, described by the quantum state vector $\vert \psi \rangle$ of the Hilbert space ${\cal H}$. The corresponding action functionals are \begin{equation}
S^T_{cl} [{\cal C}] = \int_{0 }^{T} dt [\sum_{k=1}^n p_k \dot{q}^k - H_{cl}({\cal C}) ]~~ \label{sc}
\end{equation}
for the classical system, respectively the TDSE action
\begin{equation}
S^T [\psi] = \int_0^T dt \langle \psi \vert i \hbar \partial_t - \hat{H}_0 \vert \psi \rangle \label{sq}
\end{equation}
for the quantum system. A coupling between these components can be introduced by an interaction term $\hat{h}_{\cal C}$, which is an operator on ${\cal H}$ parameterized by the classical variables ${\cal C}$. The resulting classical-quantum dynamics will be described by the action functional 
\begin{equation}
S^T [{\cal P}] = \int_0^T dt \langle {\cal P} \vert i \hbar \partial_t - \hat{H}
\vert {\cal P} \rangle \label{scq}
\end{equation}
in which $\hat{H}  =  \hat{H}_0 + \hat{h}_{\cal C}$ and 
\begin{equation}
\vert {\cal P} \rangle_{(t)} = e^{- \frac{i}{\hbar} S^t_{cl} [{\cal C}]}  \vert \psi \rangle_{ (t)} \label{p}
\end{equation}
is a product between the quantum state vector $\vert \psi \rangle$ and the classical "wave-function" 
 $e^{- \frac{i}{\hbar} S^t_{cl} [{\cal C}]}$. \\ \indent
For complex quantum systems $\hat{H}_0$ can be decomposed in a sum $\hat{H}_0 =\hat{H}_o+\hat{H}_I$ between an "orbital" term $\hat{H}_o$ and an intrinsic term $\hat{H}_I$. Therefore, it is useful to write ${\cal H}$ as a direct product ${\cal H}={\cal H}_o \otimes {\cal H}_I$ of the corresponding Hilbert spaces. This factorization is important because each term may require a specific treatment. For instance, a term $\hat{H}_o$ describing the motion of the center of mass in some external potential can be treated by the one-body TDSE. However, an intrinsic many-body Hamiltonian $\hat{H}_I$ which contains the interactions responsible for the binding energy of the system needs some mean-field approximation. Therefore, in general $\vert \psi \rangle_{ (t)}$ should be taken of the form 
\begin{equation}
\vert \psi \rangle_{ (t)}= \vert \psi_o \rangle_{ (t)} \vert \chi \rangle_{ (\tilde{x})}~~,
\end{equation}
where $\vert \psi_o \rangle_{ (t)}$ a state vector in ${\cal H}_o$ and $\vert \chi \rangle_{(\tilde{x})} \in M_I$ an element of a $2N$-dimensional symplectic trial manifold $M_I \subset {\cal H}_I$, parameterized by the real variables $\tilde{x} \equiv (x^1,x^2,...,x^{2N} )$. \\ \indent
The action functional $S^T [{\cal P}]$ depends in general on three sets of trajectories:  ${\cal C}_t$ on $M_{cl}$, $\tilde{x}_t$ on $M_I$ and $\vert \psi_o \rangle_{ (t)}$ in ${\cal H}_o$. Presuming that there is no coupling between ${\cal H}_o$ and ${\cal H}_I$ mediated by $\hat{h}_{\cal C}$, the TDVP $\delta_{{\cal C}, \psi_o, \tilde{x} }S^T[{\cal P}]=0$ provides the system of coupled equations
\begin{equation}
\dot{q}^k  =  \frac{ \partial (H_{cl} + \langle \psi \vert \hat{h}_{\cal C} \vert \psi \rangle ) }{\partial p_k} ~,~
\dot{p}_k  =  - \frac{ \partial (H_{cl} + \langle \psi \vert \hat{h}_{\cal C} \vert \psi \rangle ) }{\partial q^k}~~, k=1,n  \label{eqcc}   
\end{equation}
\begin{eqnarray}
i \hbar  \partial_t \vert \psi_o \rangle & = & (\hat{H}_o + \hat{h}_{\cal C}) \vert \psi_o \rangle  \label{eqc1} \\
 \sum_{j=1}^{2N} \dot{x}^j \omega_{jk} & = &
 \partial_k \langle \chi \vert \hat{H}_I + \hat{h}_{\cal C} \vert \chi \rangle ~~, \label{eqc4} 
\end{eqnarray}
where $\partial_k \equiv \partial / \partial x^k$ and
\begin{equation}
\omega_{jk}  = 2 \hbar {\rm Im} \langle \partial_j \chi \vert \partial_k \chi \rangle  \label{sy}
\end{equation}
are the coefficients of the symplectic form $\omega = \sum_{j<k} \omega_{jk} dx^j \wedge dx^k $ on $M_I$. 
 \\ \indent
If $M_I$ is a homogeneous space which breaks some continuous symmetry of $\hat{H}_I$ generated by an operator $\hat{P}$, ($[ \hat{P}, \hat{H}_I]=0$, $\hat{P} M_I \subset TM_I$), then it is convenient to select a submanifold of states $\vert \chi \rangle_{(\zeta, \wp)} =e^{i \zeta \hat{P} / \hbar} \vert \chi \rangle_{(0, \wp)}$, where $\vert \chi \rangle_{(0, \wp)} \equiv \vert \chi_c \rangle \in M_I$ is a continuous set of critical points (local minima) of $\hat{H}_I + c \hat{P}$, parameterized either by $c \in {\mathsf R}$ or by $\wp(c) \equiv \langle \chi_c \vert \hat{P} \vert \chi_c \rangle$. The variables $\zeta, \wp$ can be used as distinguished components of $\tilde{x}$, because  
\begin{equation}
\omega_{\zeta \wp}= -2 {\rm Re} \langle \chi_c \vert \hat{P} \partial_{\wp} \vert \chi_c \rangle=
- \partial_{\wp} \langle \chi_c \vert \hat{P} \vert \chi_c \rangle =-1
\end{equation}
and
\begin{equation}
\dot{\zeta}= - \partial_\wp \langle \chi \vert \hat{H}_I \vert \chi \rangle =c~~,~~ \dot{\wp} = \partial_\zeta 
 \langle \chi \vert \hat{H}_I \vert \chi \rangle =0 ~~.
\end{equation}
Thus, the broken symmetry of $\hat{H}_I$ is reflected by the conservation of $\wp$, and the uniform (inertial) motion of $\zeta$ with the velocity $c$. \\ \indent
It is important to remark that $S^T [{\cal P}]$ in Eq. (\ref{scq}) was defined assuming that the action functionals in Eqs. (\ref{sc}) and (\ref{sq}) are expressed in terms of the same time parameter $t$. In fact, TDSE can arise from the time-independent Schr\"odinger equation by the coupling between the quantum system and its classical environment \cite{br}. A kinematical treatment of time in quantum mechanics as a collective variable, rather than as a parameter, is presented in the Appendix.  \\[1cm]

{\bf III. The nonlinear Fokker-Planck equation} \\

The effects of a classical environment on a quantum system can be studied by using the model of bilinear coupling \cite{zw}. In this model the environment consists of independent harmonic oscillators continuously distributed in frequency with the density $\mu_\omega$. For a quantum system with no intrinsic structure (${\cal H} \equiv {\cal H}_o$) and  a single coupling operator $\hat{x}$, the interaction term has the form      
\begin{equation}
\hat{h}_{\cal C} =  \hat{x} \int_0^\infty d \omega \mu_\omega C_\omega q_\omega~~,
\end{equation}
where $C_\omega$ are constants and $q_\omega$ are the  time-dependent coordinates of the classical oscillators. Each oscillator has the Hamiltonian $(p_\omega^2  + m_\omega^2  \omega^2 q_\omega^2 )/2 m_\omega$, and Eqs.  (\ref{eqcc}) become
\begin{eqnarray} 
\dot q_\omega  =  \frac{p_\omega}{m_\omega} ~~,~~ 
\dot p_\omega  =  - m_\omega \omega^2 q_\omega - C_\omega  \langle \psi \vert  \hat{x} \vert \psi \rangle ~~.
\end{eqnarray}
The solution  $q_\omega(t)$ is a function of $Q_\psi (t) \equiv \langle \psi \vert  \hat{x} \vert \psi \rangle $, and once replaced in Eq. (\ref{eqc1}), yields the modified TDSE 
\begin{equation}
i \hbar \partial_t \vert \psi \rangle = [\hat{H}_0 + \hat{W}(t)] \vert \psi \rangle \label{eqw}
\end{equation} 
in which $\hat{ W}(t) = - \hat{ x}  \xi(t) - \hat{ x} f_\psi (t) + \hat{W}_0(t) $.  
The force functions $\xi(t)$ and $f_\psi (t)$ are defined by
$$
\xi (t)=  
- \int_0^\infty d \omega \mu_\omega C_\omega \lbrack q_\omega (0) \cos \omega t  
+ \frac{  {\dot q}_\omega (0)}{\omega} 
\sin \omega t \rbrack~~,
$$
$$
f_\psi (t) = - \int_0^t \Gamma(t-t' ) \dot{Q}_\psi (t') dt'~~,~~
\Gamma (t) = \frac{2}{\pi} \int_0^\infty d \omega J(\omega)  \frac{
\cos \omega t}{\omega} ~~,
$$
where $J(\omega) = \pi C_\omega^2 \mu_\omega/ 2 \omega m_\omega$ is the spectral density of the environment. The operator  $\hat{W}_0 (t)=-\hat{x}[Q_\psi (t) \Gamma(0) -Q_\psi (0) \Gamma(t)]$ ensures the invariance of $\hat{W}$ with respect to the renormalization of the memory function $\Gamma$ by an additive constant, and vanishes when $\Gamma (t) \sim \delta (t)$ or $Q_\psi(t)=0$. \\ \indent
The modified TDSE (\ref{eqw}) describes the quantum Brownian motion when the initial values $(q_\omega (0), p_\omega (0))$ are distributed within a statistical ensemble ${\cal E}$ at the temperature $T$ \cite{zw, cl}. In this case 
\begin{equation}
<<  \xi(t) >>_{\cal E} =0 ~~,~~ <<  \xi(t) \xi(t') >>_{\cal E} =  k_B T \Gamma(t-t') ~~
\end{equation}
where $ <<  * >>_{\cal E}$ denotes the average over ${\cal E}$. In practice, $<< * >>_{\cal E}$ is replaced by the average $\ll * \gg$, over an ensemble ${\cal B}$ of $N_B$ Brownian trajectories $\{ \vert \psi \rangle^k_{ (t)}, k=1,N_B ; \vert \psi \rangle^k_{ (0)}= \vert \psi_0 \rangle \}$, generated from the same initial condition $\vert \psi_0 \rangle$. Thus, $\ll \xi(t) \gg =0$,
\begin{equation}
\ll \xi(t) \xi(t') \gg = k_B T \Gamma(t-t') ~~,  \label{FDT}
\end{equation}
and $\xi(t)$ behaves like a stochastic force with zero mean, related to the memory function $\Gamma(t)$ by the fluctuation-dissipation theorem. Assuming that in this case $\hat{W}_0 (t)=0$, then Eq. (\ref{eqw}) becomes a  stochastic, nonlinear Schr\"odinger-Langevin equation\footnote{Despite the similar name, this equation is not the same as the one in  \cite{kost,raz}.}  
\begin{equation}
i \hbar \partial_t \vert \psi \rangle = [ \hat{H}_0  
- \hat{ x} \xi(t) - \hat{ x} f_\psi (t)  ] \vert \psi \rangle~~. \label{sle}
\end{equation}
Here $- \hat{ x} \xi(t)$ describes the noise-induced transitions, while $- \hat{ x}f_\psi (t)$ corresponds to the phenomenological dissipative term used in \cite{alb} to obtain Ehrenfest equations with friction. \\ \indent
A common example is the Ohmic environment, when $J(\omega) = \gamma \omega $ and $\Gamma(t) = 2 \gamma \delta(t)$, by $\gamma$ denoting the friction coefficient. 
The coupling term $\hat{h}_e = -e {\bf p} \cdot {\bf A}/mc$ between a quantum charged particle and a thermal radiation field ${\bf A}(t)$ corresponds to $J_e (\omega) = 2 \pi^2 e^2 \omega u_\omega /3  k_B T$, where $u_\omega$ is the spectral energy density of the thermal radiation. The memory function $\Gamma_e (t)$ vanishes when $T=0$, but at high temperatures it becomes $\Gamma_e (t)  = -  4 e^2 \ddot{\delta}(t) / 3 c^3$ \cite{pha}. A similar result can be expected also for a quantum particle coupled to a thermal gravitational radiation field by $\hat{h}_g=  {\bf p} \cdot {\bf A}_g /c$, where ${\bf A}_g$ is the gravitomagnetic vector potential\footnote{ $g_{\mu \nu} = g_{\mu \nu}^{(0)}+h_{\mu \nu}$, $g_{\mu \nu}^{(0)}={\rm diag}[1,-1,-1,-1]$,  $h_{00}= 2 \Phi_g/c^2$, $h_{0i}= ({\bf A}_g)_i/c^2$}  \cite{bm, kw}. 
\\ \indent
For the quantum system described by Eq. (\ref{sle}), the value of the observable associated to an operator $\hat{K}$ should be defined by   
\begin{equation}
\ll \langle \psi \vert \hat{K} \vert \psi \rangle \gg = Tr( \hat{\rho}_{av}(t) \hat{K}) \label{exv}
\end{equation}
where $\hat{\rho}_{av} \equiv \ll \hat{\rho} \gg $ is the average density matrix. Along each Brownian trajectory $\vert \psi \rangle_{ (t)} \in {\cal B}$,  $\hat{\rho}= \vert \psi \rangle \langle \psi \vert$ satisfies the nonlinear Liouville-Langevin equation  
\begin{equation}
i \hbar \partial_t \hat{\rho} = [ \hat{H}_0  
- \hat{ x} \xi(t) - \hat{ x} f_\rho (t)  , \hat{ \rho}]~~, \label{lle}
\end{equation} 
where  $f_\rho (t) = - \int_0^t \Gamma(t-t' ) \dot{Q}_\rho (t') dt'$ and $Q_\rho= Tr(\hat{x} \hat{\rho} )$. Thus $\rho_{av}(t)$ can be calculated in principle by taking the average over an ensemble of trajectories $\hat{\rho}(t)$  generated by (\ref{lle}) with the same initial condition $\hat{\rho}(0)=\hat{\rho}_0$. However, this procedure is time-consuming, and unless a large number of observables is needed, it is more convenient to calculate $\ll \langle \psi \vert \hat{K} \vert \psi \rangle \gg$ for each $\hat{K}$ of interest while solving Eq. (\ref{sle}). An alternative to the ensemble average is to integrate the transport equation satisfied by $\hat{\rho}_{av}$. For short times, this equation can be obtained by starting from the expansion 
\begin{equation}
\hat{\rho}(t+ s) = \hat{\rho} (t) + ( \partial_t \hat{\rho}) s + (\frac{ \partial^2 \hat{\rho}}{\partial t^2}  ) \frac{s^2}{2!}
\end{equation}
where $s$ is a small time step. The ensemble average  
\begin{equation}
\hat{\rho}_{av} (t+ s) = \hat{\rho}_{av}(t) + \ll \partial_t \hat{\rho}\gg s + \ll \frac{  \partial^2 \hat{\rho}}{\partial t^2}\gg  \frac{s^2}{2!}
\end{equation}
shows that $\hat{\rho}_{av} (t)$ is the solution of the equation 
\begin{equation}
i \hbar \partial_t \hat{\rho}_{av} (t) =  i \hbar \ll \partial_t \hat{\rho} \gg + \frac{i \hbar}{2!}  \lim_{s \rightarrow 0} \ll \frac{  \partial^2 \hat{\rho}}{\partial t^2}\gg  s \label{av} ~~.
\end{equation}
The coefficients $ \ll \partial_t \hat{\rho}\gg$ and $\ll\partial_t^2 \hat{\rho}\gg$ can be estimated by using Eq. (\ref{lle}), and depend on the explicit form of the memory function $\Gamma(t)$.  
Let us consider the case of linear friction, when  
\begin{equation}
f_\rho = - \gamma Tr(\hat{x} \partial_t \hat{\rho} ) = \frac{i}{\hbar} \gamma Tr([\hat{x}, \hat{H}_0] \hat{\rho})~~.
\end{equation}
Presuming that the equalities 
\begin{equation}
\ll \xi \hat{\rho} \gg = \ll \xi \gg \ll \hat{\rho} \gg   ~~,~~\ll \xi^2 \hat{\rho} \gg = \ll \xi^2\gg \ll \hat{\rho} \gg \label{frx}
\end{equation}
and
\begin{equation}
\ll f_\rho \hat{\rho} \gg = \ll f_\rho \gg \ll \hat{\rho} \gg
\end{equation}
which are satisfied at $t=0$ remain valid at any time $t$, we obtain 
\begin{equation}
i \hbar \ll \partial_t \hat{\rho}\gg = [ \hat{H}_0 - i \hat{ x} \frac{\gamma}{\hbar} Tr([\hat{x}, \hat{H}_0] \hat{\rho}_{av}) , \hat{ \rho}_{av}]~~,  \label{av1}
\end{equation} 
and 
\begin{equation}
\lim_{s \rightarrow 0} \ll \frac{  \partial^2 \hat{\rho}}{\partial t^2}\gg  s =
 - \frac{[\hat{x}, [\hat{x}, \hat{\rho}_{av} ]]}{\hbar^2} \lim_{s \rightarrow 0}  \ll \xi^2 \gg s~~. \label{av2}   
\end{equation}
Here $\lim_{s \rightarrow 0} \ll \xi^2 \gg s=2 k_B T \gamma$, because $s$ can be taken as the step of a time-grid $\{t_k=ks , k=1,2,3,... \}$ in which $\xi(t)$ is normalized according to
\begin{equation}
\ll \xi (t_j ) \xi (t_k )\gg =2 k_B T \gamma \frac{\delta_{t_j t_k}}{s}  ~~.
\end{equation}
Replacing now (\ref{av1}) and (\ref{av2}) in (\ref{av}) we obtain the transport (Fokker-Planck) equation 
\begin{equation}
i \hbar \partial_t \hat{\rho}_{av} = [\hat{H}_0 -i \hat{ x} \frac{\gamma}{\hbar} Tr([\hat{x}, \hat{H}_0] \hat{\rho}_{av}) , \hat{\rho}_{av}]
- \frac{ i \gamma k_B T}{ \hbar} [\hat{x}, [\hat{x}, \hat{\rho}_{av} ]] ~~. \label{fpe}
\end{equation} 
\\

{\bf IV. Applications to the quantum two-state system} \\

Quantum systems having two states, $\vert 0\rangle$ and $\vert 1\rangle$, interacting with the environment, are used to describe spin relaxation, tunneling in double-well potentials, or optical transitions between internal atomic levels.  
\\ \indent
The two quantum states will be chosen here as eigenstates of the Hamiltonian $\hat{H}_0$, separated in energy by $\Delta \equiv \hbar \omega_0$, so that $\hat{H}_0$ and $\hat{x}$ are 
\begin{equation}
\hat{H}_0 = \frac{\Delta}{2}(\vert 1\rangle \langle 1 \vert -
\vert 0 \rangle \langle 0 \vert)
~~,~~ \hat{x}= Q i(\vert 0\rangle \langle 1 \vert - \vert 1 \rangle \langle 0 \vert)~~.
\end{equation}
Using the representation
$$ 
\vert 0\rangle = \left[ \begin{array}{c} 0 \\ 1 \end{array} \right] ~~,~~
\vert 1 \rangle =\left[ \begin{array}{c} 1 \\ 0 \end{array} \right] 
$$
the density matrix $\hat{\rho}$ has the general form 
\begin{equation} 
\hat{\rho} = \frac{1}{2} \left[ \begin {array}{cc} 1+b_1 & b_2-ib_3 \\ b_2+ib_3 & 1-b_1
\end{array} \right]
\end{equation}
parameterized by the polarization (Bloch) vector ${\bf b}=(b_1,b_2,b_3)$.
\\ \indent
In terms of ${\bf b}$ and its average ${\bf P} = \ll {\bf b} \gg$, Eqs. (\ref{lle}) and (\ref{fpe}) become  
\begin{eqnarray}
 \dot{ b}_1 & = &2 Q \xi b_2 / \hbar - A_0 b_2^2 \nonumber \\
 \dot{ b}_2 & = & - \omega_0 b_3 - 2 Q \xi b_1/ \hbar + A_0 b_1b_2  \label{sle1} \\
 \dot{ b}_3 & = & \omega_0 b_2  \nonumber
\end{eqnarray}
respectively
\begin{eqnarray}
\dot{ P}_1 & = & - 2 \lambda P_1  - A_0 P_2^2  \nonumber \\
\dot{ P}_2 & = & - \omega_0 P_3 -2 \lambda P_2 + A_0 P_1P_2  \label{fpe1} \\
\dot{ P}_3 & = & \omega_0 P_2 ~~,           \nonumber
\end{eqnarray}
where $\lambda = 2 \gamma Q^2 k_B T / \hbar^2$ is the rate of the noise-induced transitions between the two levels and $A_0= 2 \gamma Q^2 \Delta / \hbar^2$. These equations could be applied, for instance, to the density matrix of the molecular electron considered in \cite{boa}, when the collective coordinate is practically fixed at the ground-state value, and only its conjugate momentum has thermal fluctuations\footnote{ $(b_1,b_2,b_3)$ correspond, respectively, to the variables $(z,x,y)$ used in \cite{boa}.}.  
\noindent
\begin{figure}
\begin{picture}(100,180)(0,0)
\end{picture}
\vskip.2cm
Figure 1. $P_1$ as a function of time when $\vert \psi_0 \rangle = \cos \theta \vert 1 \rangle + i \sin \theta \vert 0 \rangle$, for $\theta = \pi /4$ (solid), $\pi /12$ (dash) and $\pi /50$ (*).  
\end{figure}
\\ \indent
Let us assume that initially the system is prepared in a pure state $\vert \psi_0 \rangle= \cos \theta \vert 1 \rangle + i \sin \theta \vert 0 \rangle$, corresponding to the polarization vector ${\bf b}_0=(\cos 2 \theta, 0, \sin 2 \theta)$. 
If $A_0 \approx 0$ but  $\lambda >0$, then when $\vert \psi_0 \rangle = (-i)^{\sigma -1}  \vert \sigma \rangle$, $\sigma =0,1$ is one of the eigenstates ($\theta= \pi/2$ or $0$), Eq. (\ref{fpe1}) has the solution    
\begin{equation}
{\bf P} (t)  =   e^{-2 \lambda t}((-1)^{\sigma+1},0,0)~~.  \label{psts} \\
\end{equation}
For the nonstationary superposition $\vert \psi_0 \rangle  = (\vert 1 \rangle + i \vert 0 \rangle)/\sqrt{2}$, ($\theta = \pi/4$), in the same approximation the solution is 
\begin{equation}
{\bf P}(t)  =  e^{- \lambda t} (0, - \frac{\omega_0}{\omega} \sin \omega t ,
\cos \omega t + \frac{\lambda}{\omega} \sin \omega t) \label{pnsts} 
\end{equation}
where $\omega = \sqrt{\omega_0^2 - \lambda^2}$. 
In both cases, asymptotically $\vert  {\bf P} \vert \rightarrow 0$, and the initial state  evolves into an incoherent mixture between $\vert 0\rangle$ and $\vert 1\rangle$. The exact solutions (\ref{psts}) and (\ref{pnsts}) provide a very good  fit for the ensemble average generated by integrating numerically (\ref{sle1}) with $k_B T \approx 4 \Delta$, $A_0 =2 \cdot 10^{-4} \omega_0$, and the same initial conditions \cite{pha}.  \\ \indent
At zero temperature the thermal noise vanishes ($\lambda=0$) and the coupling to the classical environment produces only a dissipative dynamics. In this case $\hat{\rho}_{av} \equiv \hat{\rho}$, ${\bf P} \equiv {\bf b}$, and both equations, (\ref{sle1}) and (\ref{fpe1}), reduce  to 
\begin{eqnarray}
\dot{ P}_1 & = &   - A_0 P_2^2  \nonumber \\
\dot{ P}_2 & = & - \omega_0 P_3 + A_0 P_1P_2  \label{pfr} \\
\dot{ P}_3 & = & \omega_0 P_2 ~~,           \nonumber
\end{eqnarray}
\noindent
\begin{figure}
\begin{picture}(100,180)(0,0)
\end{picture}
\vskip.2cm
Figure 2. $P_k$ (solid) and  $\ll b_k \gg$ (*), $k=1,2,3$, given by Eqs. (\ref{fpe1}) and (\ref{sle1}) for the initial state $\vert  \psi_0 \rangle  = (\vert 1 \rangle + i \vert 0 \rangle)/\sqrt{2}$, as a function of time.  
\end{figure}
\noindent
It is easy to see that ${\bf P} (t) =((-1)^{\sigma+1},0,0)$, $\sigma=0,1$ are constant solutions of (\ref{pfr}), so that the system may stay on a stationary level an indefinite time. However, between the two only $\vert 0 \rangle$ is stable, because any admixture of $\vert 0 \rangle$ to $\vert 1  \rangle$, no matter how small, will induce its irreversible decay. The  short-time behaviour of the component $P_1$ when $A_0=1$, $\omega_0=5$, $\hbar=1$, is illustrated in Figure 1. For any initial condition $\vert \psi_0 \rangle \ne \vert 0 \rangle, \vert 1 \rangle$, in the asymptotic region $t \gg t_a$, defined by the moment $t_a$ when $P_1(t_a) \approx -1$, $P_2(t_a) =0$, and $P_3 (t_a) =a \ne 0$, the system described by (\ref{pfr}) approaches the ground state according to  
\begin{eqnarray}
P_1(t) & = &-1 + a^2 e^{- A_0 (t-t_a)} \frac{1 - \cos \varphi \cos (2 \omega (t- t_a)+ \varphi)}{ 2 \sin^2 \varphi} \\
P_2(t) & = & -  a e^{-A_0(t-t_a)/2} \frac{ \sin \omega (t -t_a)}{\sin \varphi} \\
P_3(t) & = &  a e^{-A_0(t-t_a)/2} [ \cos \omega (t-t_a) + \frac{A_0}{2 \omega}  \sin \omega (t-t_a)] 
\end{eqnarray}
where $\omega = \sqrt{\omega_0^2 - A^2_0/4}$ and $\sin \varphi = \omega/ \omega_0$. Therefore, after a transitory re\-gime the energy $ \Delta P_1/2$ decreases exponentially due to the factor $e^{- A_0 t}$, the same as in the spontaneous decay found when the environment is quantized.  \\ \indent
A comparison between ${\bf P}$ and  $\ll {\bf b} \gg$
for the initial state $\vert \psi_0 \rangle  = (\vert 1 \rangle + i \vert 0 \rangle)/\sqrt{2}$ when both terms, dissipation and noise, are significant is presented in Figure 2. All parameters are the same specified above for Figure 1, excepting for the temperature $T$ which is  $0.02 \Delta / k_B$ in Fig. 2A  and $2 \Delta /k_B$ in Fig. 2B. The results show that $P_2(t)$ and $P_3 (t)$ provided by (\ref{fpe1}) practically coincide with $\ll b_2(t) \gg$ and $\ll b_3(t) \gg$
calculated using an ensemble of $N_B$ solutions ($N_B$=500 in Fig 2A and 2000 in Fig. 2B) of Eq. (\ref{sle1}). Significant differences appear however between the asymptotic limits of $P_1(t)$ and $\ll b_1(t) \gg$. 
Thus, by contrast to $\ll b_1(t) \gg$, which shows thermalization at an equilibrium value $b_{1eq} \equiv\ll b_1(t \rightarrow \infty ) \gg $ increasing with $T$ (Figure 3), Eq. (\ref{fpe1}) yields $P_1(t \rightarrow \infty)=0$ and complete decoherence. \\
\noindent
\begin{figure}
\begin{picture}(100,180)(0,0)
\end{picture}
\vskip.2cm
Figure 3. $b_{1eq}$ (*) compared to the analytic expressions $f_1 = - A_0/(2 \lambda + A_0)$ (solid) and $f_2 = - A_0/(4 \lambda + A_0)$ (dash)  as a function of $k_BT/ \Delta$.
\end{figure}

{\bf V. Summary and conclusions} \\

The time-dependent variational principle provides a general framework to derive dynamical equations in classical and quantum mechanics.  In Section II it was shown that the evolution of a system consisting of interacting classical, quasiclassical and quantum components can be described by a time-dependent variational principle for the TDSE action. The classical variables appear both in the phase-factor of the state vectors, and in the Hamilton operator of the quantum subsystem. The separation of a quasiclassical sector in the Hilbert space, considered usually part of an approximation, can also have a fundamental relevance in the treatment of time as a dynamical variable. This aspect is discussed in Appendix, assuming that the time wave-functions are coherent states.  
\\ \indent
A bilinear coupling between a system of classical oscillators (environment) and a quantum particle leads to an effective Hamiltonian containing an external force term and  the retarded backreaction of the environment. If the oscillators belong to an ensemble in thermal equilibrium, these terms become the noise, respectively the friction, changing TDSE into a nonlinear Schr\"odinger-Langevin equation.  The calculation of the observables for a quantum Brownian particle described by this equation was discussed in Section III. The environment is specified by its spectral density function, given explicitly for a classical thermal electromagnetic field and for the simple case of Ohmic dissipation. Measurable quantities, such as decoherence rates, can be extracted either by statistical average over an ensemble of quantum expectation values, or from the expectation value given by the average density matrix. It is shown that for a classical Ohmic environment, in certain conditions this matrix can be obtained by integrating a nonlinear quantum Fokker-Planck equation. 
\\ \indent 
Applications to a generic two-state system are presented in Section IV, considering both stationary and nonstationary initial states. Without friction ($A_0=0$, $\lambda >0$), the average polarization vector is explicitly obtained as a function of time, temperature, and the model parameters (Eqs. \ref{psts}, \ref{pnsts}). This function is close to the one calculated numerically from Eq. (\ref{sle1}) by ensemble average \cite{pha}, showing that the factorizations presumed in Eq. (\ref{frx}) are reliable. The case of zero temperature ($A_0 >0$, $\lambda=0$) is singular because the classical environment produces an irreversible, spontaneous decay (Figure 1) only if the initial state is, even to a small degree, nonstationary. Also when friction is comparable, or stronger than noise ($A_0 \geq \lambda >0$), Eqs. (\ref{lle}) and (\ref{fpe}) are equivalent only during a fraction of the thermalization time (Figure 2). Therefore, to reproduce the low-temperature behaviour presented in Figure 3, Eq. (\ref{fpe}) should contain additional terms. \\ \indent
It is important to remark that for most physical systems, the assumption of a classical environment used to derive Eqs. (\ref{lle}, \ref{fpe}) becomes unrealistic at very low temperature. An exception, worth of further consideration, could be provided by the interesting open problem of the interaction between matter and thermal gravitational radiation.   \\

{\bf Appendix} \\[.5cm] \indent
The phase-space $M_{cl}$ of a classical system can be extended to $M_{cl}^e= M_{cl} \times T^*{\mathsf R}$ by two canonical coordinates $q^0= c t$, $p_0 = - E /c$ associated to the time and energy variables $(t, E)$ \cite{et}. The trajectories ${\cal C}^e$ in this extended phase-space are parameterized by a new variable, $u$ ("universal" time), and the action functional takes the form  
\begin{equation}
S^U_e [{\cal C}^e] = \int_{0 }^{U} du [\sum_{k=0}^n p_k d_uq^k - H_{cl}^e({\cal C}^e) ]~~,
\end{equation}
where $H^e_{cl} ({\cal C}^e)= H_{cl}({\cal C})+cp_0$ and $d_u \equiv d/du$. \\ \indent
Similarly, the extended Hilbert space ${\cal H}^e$ for a scalar quantum particle described by the space-time coordinates $({\bf R}, T)$  should be defined by a set of state vectors $ \psi^e $ integrable over all variables. Thus, ${\cal H}^e$ is a direct product ${\cal H}^e={\cal H}_o \otimes {\cal H}_I$ between the spaces
\begin{equation}
{\cal H}_o  =   \{ \psi ({\bf R}) , \int d^3 R ~\vert \psi ({\bf R})  \vert^2 =1 \}
\end{equation}
and
\begin{equation}
{\cal H}_I  =  \{ \chi (T), \int d T ~\vert \chi (T) \vert^2 =1 \} ~~.
\end{equation}
In this representation, the coordinate and momentum operators on ${\cal H}_I$ can be defined by the relations $\hat{T}= T$, $\hat{\Pi}=- i \hbar \partial/c \partial T$.    \\ \indent
For a quantum particle interacting with a classical environment we can presume an extended action functional of the form   
\begin{equation}
S^U [{\cal P}^e] = \int_0^U du \langle {\cal P}^e \vert i \hbar \partial_u - \hat{H}^e
\vert {\cal P}^e \rangle
\end{equation}
where 
\begin{equation}
 {\cal P}^e  (u) = e^{- \frac{i}{\hbar} S^u_e [{\cal C}^e]} \psi^e  (u) 
\end{equation}
and $\hat{H}^e  =  \hat{H}_0 + c \hat{\Pi} + \hat{h}_{\cal C}$. The second term  $c \hat{\Pi} $ corresponds to $cp_0$ from the classical extended Hamiltonian. However, it can be seen also as the symmetry-restoring term required by a physical state $\psi^e$ localized in time. Therefore, $\hat{H}^e$ appears as the effective, one-particle term, of a "mean-field" approximation for a more fundamental Hamiltonian containing the strong many-body interaction which confines the particles in time around present. This localization can be taken into account by a quasiclassical treatment of time within the manifold $M_I \subset {\cal H}_I$ of the Gaussian wave-packets $\chi_z$, 
\begin{equation}
 \chi_z = e^{z \hat{B}^\dagger - z^* \hat{B} } \chi_0 ~~,~~ 
z = ( \Omega \tau - i \epsilon / \hbar \Omega  ) / \sqrt{2}
\end{equation}
parameterized as coherent states, by the real variables $\tilde{x} \equiv ( \tau , \epsilon)$.
Here $\hat{B}, \hat{B}^\dagger$ are the Dirac-Fock operators,
$$
\hat{B} = \frac{1}{\sqrt{2}} (\Omega T + \frac{1}{\Omega} \frac{\partial}{\partial T} )~~,~~ [\hat{B}, \hat{B}^\dagger]=1  
$$ 
and
\begin{equation}
\chi_0( T) = \sqrt{ \frac{ \Omega}{ \sqrt{ \pi}}}  e^{- \Omega^2 T^2/2}
\end{equation}
is the reference element defined by $\hat{B} \chi_0 =0$. The width $\sim 1/\Omega$ of these wave-packets can be seen as the time-interval defining the present. For this choice the coefficients of the symplectic form on $M_I$ are $\omega_{\epsilon \tau} = - \omega_{\tau \epsilon} = 1$. \\ \indent
The application of the TDVP $\delta S^U[{\cal P}^e]=0$ to the variations of ${\cal C}^e$ and $ \psi^e = \psi \chi_z $ in $M^e_{cl} \times {\cal H}_o \times M_I$ provides the system of coupled equations
\begin{equation}
d_uq^k  =  \frac{ \partial (H_{cl} + \langle \psi^e \vert \hat{h}_{\cal C} \vert \psi^e \rangle ) }{\partial p_k} ~,~
d_up_k  =  - \frac{ \partial (H_{cl} + \langle \psi^e \vert \hat{h}_{\cal C} \vert \psi^e \rangle ) }{\partial q^k}~,k=1,n \label{dc}   
\end{equation}
\begin{equation}
d_u q^0  =  c ~~,~~ 
d_u p_0  =  - \frac{ \partial (H_{cl} + \langle \psi^e \vert \hat{h}_{\cal C} \vert \psi^e \rangle ) }{\partial q^0} \label{tc}
\end{equation}
\begin{equation}
i \hbar  \partial_u  \psi   =  (\hat{H}_0 + \hat{h}_{\cal C})  \psi   \label{dq} 
\end{equation}
\begin{equation}
d_u \tau  =  - \partial_\epsilon \langle \chi_z \vert c \hat{\Pi} \vert \chi_z \rangle ~~,~~
d_u \epsilon  =   \partial_\tau \langle \psi^e \vert \hat{H}_0 + \hat{h}_{\cal C} \vert \psi^e \rangle
\label{tq}
\end{equation}
In Eq. (\ref{tq}), $\langle \chi_z \vert \hat{\Pi} \vert \chi_z \rangle = - \epsilon /c$ yields 
$d_u \tau =1$, similar to $d_u q^0 =c$ from Eq. (\ref{tc}). Therefore, the centroid 
$\tau = \langle \chi_z \vert  \hat{T} \vert \chi_z \rangle $ of the Gaussian $\chi_z$, and the classical time coordinate $t=q^0/c$ perform uniform translations along the $u$-axis with the same speed. By choosing $t = \tau = u$, Eqs. (\ref{dc}) and (\ref{dq}) become the classical Hamilton equations and TDSE, respectively. \\ \indent
Within this framework, the common time shared by classical and quantum systems resembles the collective variables used in the many-body theory, and might be a result of the correlations induced by a strong interaction between the quantum states in the extended Hilbert space.

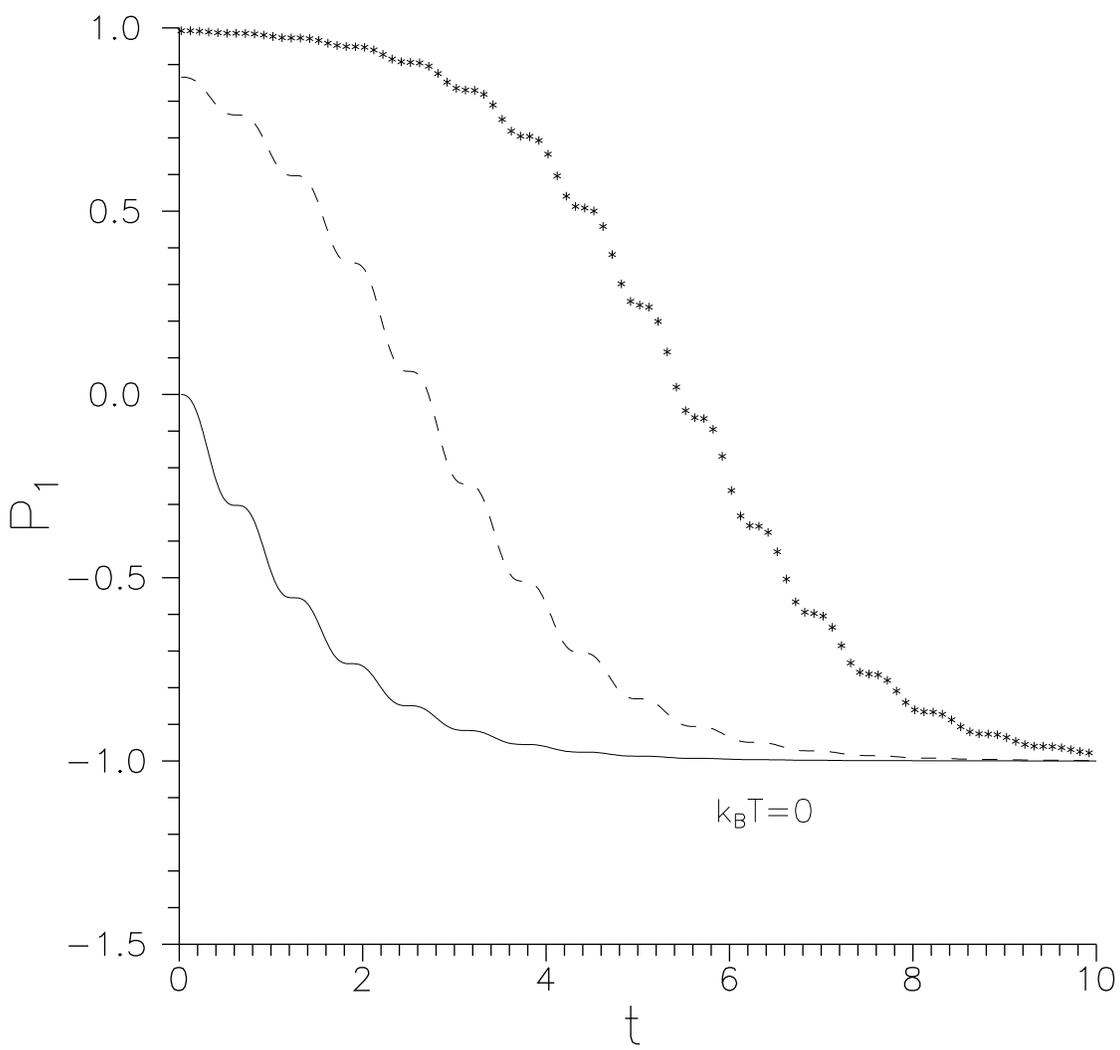
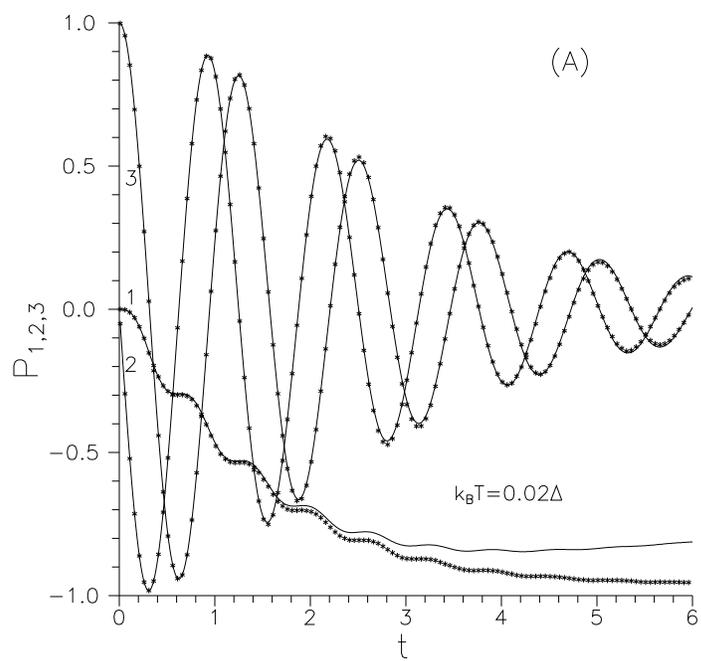
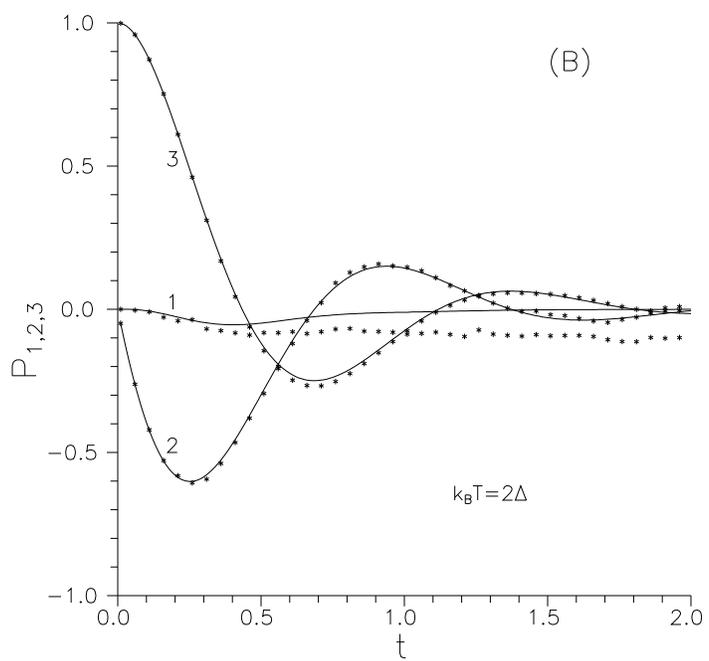
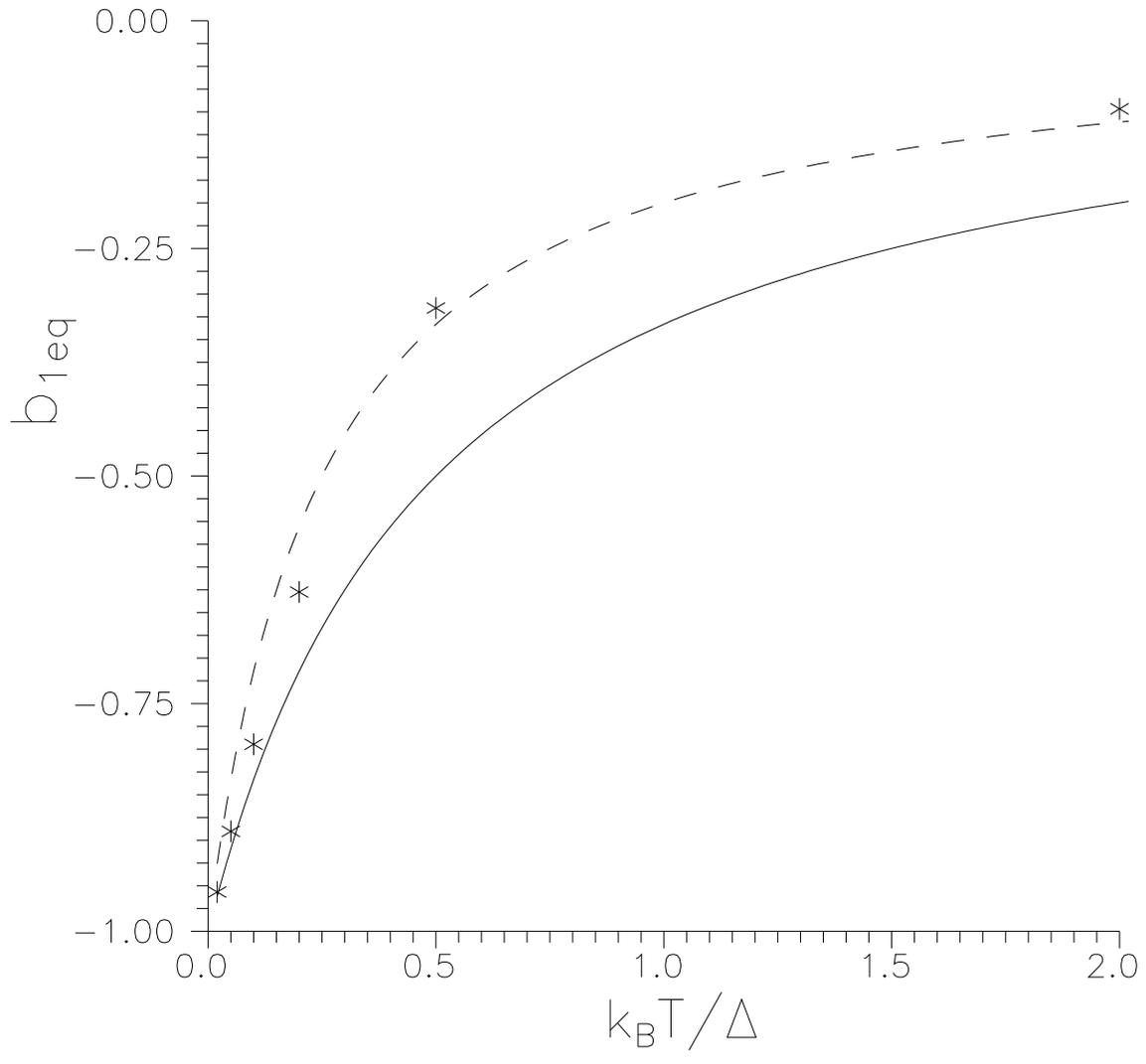
\end{document}